%% file: main.tex
\documentclass[reprint,
superscriptaddress,
 amsmath,amssymb,
 aps,prd,longbibliography]{revtex4-2}

\usepackage[table,x11names,dvipsnames,table]{xcolor}
\usepackage{siunitx}
\usepackage{hyperref}
\usepackage{graphicx}
\usepackage{dcolumn}
\usepackage{bm}

\newcommand{\uCCCoh}{\nu_\mu\text{CC-Coh}\pi}

\begin{document}
\preprint{FERMILAB-PUB-24-0581-PPD}

\title{Search for a Hidden Sector Scalar from Kaon Decay in the Di-Muon Final State at ICARUS}
\input{author_institutions}
\collaboration{ICARUS Collaboration}

\date{\today}

\begin{abstract}
\input{abstract}
\end{abstract}
\maketitle


\section{Introduction}
\input{Introduction}

\input{Models}

\section{Simulation and Event Selection}
\label{sec:selection}
\input{SignalBox}

\section{Systematic Uncertainties and Statistical Procedure}
\input{Systematics}

\input{SignalBoxStat}

\section{Results}
\input{Results}

\section{Conclusion}
\input{Conclusion}

\begin{acknowledgements}
\input{acknowledgements}
\end{acknowledgements}


\appendix
\section{$\uCCCoh$ Modeling}
\label{app:uCCCoh}
\input{cohpimodel}

\bibliography{main}

\end{document}

%% file: author_institutions.tex
\affiliation{Brookhaven National Laboratory, Upton, NY 11973, USA}
\affiliation{CBPF, Centro Brasileiro de Pesquisas F\'isicas, Rio de Janeiro, RJ 22290-180, Brazil}
\affiliation{CERN, European Organization for Nuclear Research 1211 Gen\`eve 23, Switzerland, CERN}
\affiliation{University of Chicago, Chicago, IL 60637, USA}
\affiliation{CIEMAT, Centro de Investigaciones Energ\'eticas, Medioambientales y Tecnol\'ogicas, Madrid E-28040, Spain}
\affiliation{Centro de Investigacion y de Estudios Avanzados del IPN (Cinvestav), Mexico City}
\affiliation{Columbia University, New York, NY 10027, USA}
\affiliation{Colorado State University, Fort Collins, CO 80523, USA}
\affiliation{Fermi National Accelerator Laboratory, Batavia, IL 60510, USA}
\affiliation{University of Granada, E-18071 Granada, Spain}
\affiliation{University of Houston, Houston, TX 77204, USA}
\affiliation{INFN sezione di Bologna University, Bologna, Italy}
\affiliation{INFN Sezione di Catania and University, Catania, Italy}
\affiliation{INFN Sezione di Genova and University, Genova, Italy}
\affiliation{INFN GSSI, L'Aquila, Italy}
\affiliation{INFN LNGS, Assergi,  Italy}
\affiliation{INFN LNS, Catania, Italy}
\affiliation{INFN Sezione di Milano, Milano, Italy}
\affiliation{INFN Sezione di Milano Bicocca, Milano, Italy}
\affiliation{INFN Sezione di Napoli, Napoli, Italy}
\affiliation{INFN Sezione di Padova and University, Padova, Italy}
\affiliation{INFN Sezione di Pavia and University, Pavia, Italy}
\affiliation{INFN Sezione di Pisa, Pisa, Italy}
\affiliation{Lancaster University, Lancaster LA1 4YW, United Kingdom}
\affiliation{Physical Research Laboratory, Ahmedabad, India }
\affiliation{University of Pittsburgh, Pittsburgh, PA 15260, USA}
\affiliation{University of Rochester, Rochester, NY 14627, USA}
\affiliation{University of Sheffield, Department of Physics and Astronomy, Sheffield S3 7RH, United Kingdom}
\affiliation{SLAC National Accelerator Laboratory, Menlo Park, CA 94025, USA}
\affiliation{Southern Methodist University, Dallas, TX 75275, USA}
\affiliation{University of Texas at Arlington, Arlington, TX 76019, USA}
\affiliation{Tufts University, Medford, MA 02155, USA}
\affiliation{Virginia Tech, Blacksburg, VA 24060, USA}
\affiliation{York University, York, Toronto M3J 1P3, Canada}
\author{F. Abd Alrahman}
\affiliation{University of Houston, Houston, TX 77204, USA}
\author{P. Abratenko}
\affiliation{Tufts University, Medford, MA 02155, USA}
\author{N. Abrego-Martinez}
\affiliation{Centro de Investigacion y de Estudios Avanzados del IPN (Cinvestav), Mexico City}
\author{A. Aduszkiewicz}
\affiliation{University of Houston, Houston, TX 77204, USA}
\author{F. Akbar}
\affiliation{University of Rochester, Rochester, NY 14627, USA}
\author{L. Aliaga Soplin}
\affiliation{University of Texas at Arlington, Arlington, TX 76019, USA}
\author{R. Alvarez Garrote}
\affiliation{CIEMAT, Centro de Investigaciones Energ\'eticas, Medioambientales y Tecnol\'ogicas, Madrid E-28040, Spain}
\author{M. Artero Pons}
\affiliation{INFN Sezione di Padova and University, Padova, Italy}
\author{J. Asaadi}
\affiliation{University of Texas at Arlington, Arlington, TX 76019, USA}
\author{W. F. Badgett}
\affiliation{Fermi National Accelerator Laboratory, Batavia, IL 60510, USA}
% ...
\
\altaffiliation[On leave of absence from ]{INFN Pavia}
\affiliation{Fermi National Accelerator Laboratory, Batavia, IL 60510, USA}
\author{A. Montanari}
\affiliation{INFN sezione di Bologna University, Bologna, Italy}
\author{M. Mooney}
\affiliation{Colorado State University, Fort Collins, CO 80523, USA}
\author{G. Moreno-Granados}
\affiliation{Virginia Tech, Blacksburg, VA 24060, USA}
\author{J. Mueller}
\affiliation{Colorado State University, Fort Collins, CO 80523, USA}
\author{M. Murphy}
\affiliation{Virginia Tech, Blacksburg, VA 24060, USA}
\author{D. Naples}
\affiliation{University of Pittsburgh, Pittsburgh, PA 15260, USA}
\author{V.C.L Nguyen}
\affiliation{University of Sheffield, Department of Physics and Astronomy, Sheffield S3 7RH, United Kingdom}
\author{S. Palestini}
\affiliation{CERN, European Organization for Nuclear Research 1211 Gen\`eve 23, Switzerland, CERN}
\author{M. Pallavicini}
\affiliation{INFN Sezione di Genova and University, Genova, Italy}
\author{V. Paolone}
\affiliation{University of Pittsburgh, Pittsburgh, PA 15260, USA}
\author{R. Papaleo}
\affiliation{INFN LNS, Catania, Italy}
\author{L. Pasqualini}
\affiliation{INFN sezione di Bologna University, Bologna, Italy}
\author{L. Patrizii}
\affiliation{INFN sezione di Bologna University, Bologna, Italy}
\author{L. Paudel}
\affiliation{Colorado State University, Fort Collins, CO 80523, USA}
\author{L. Pelegrina-Guti\'errez}
\affiliation{University of Granada, E-18071 Granada, Spain}
\author{G. Petrillo}
\affiliation{SLAC National Accelerator Laboratory, Menlo Park, CA 94025, USA}
\author{C. Petta}
\affiliation{INFN Sezione di Catania and University, Catania, Italy}
\author{V. Pia}
\affiliation{INFN sezione di Bologna University, Bologna, Italy}
\author{F. Pietropaolo}
\altaffiliation[On leave of absence from ]{INFN Padova}
\affiliation{CERN, European Organization for Nuclear Research 1211 Gen\`eve 23, Switzerland, CERN}
\author{F. Poppi}
\affiliation{INFN sezione di Bologna University, Bologna, Italy}
\author{M. Pozzato}
\affiliation{INFN sezione di Bologna University, Bologna, Italy}
\author{G. Putnam}
\email{gputnam@fnal.gov}
\affiliation{Fermi National Accelerator Laboratory, Batavia, IL 60510, USA}
\author{X. Qian}
\affiliation{Brookhaven National Laboratory, Upton, NY 11973, USA}
\author{A. Rappoldi}
\affiliation{INFN Sezione di Pavia and University, Pavia, Italy}
\author{G. L. Raselli}
\affiliation{INFN Sezione di Pavia and University, Pavia, Italy}
\author{S. Repetto}
\affiliation{INFN Sezione di Genova and University, Genova, Italy}
\author{F. Resnati}
\affiliation{CERN, European Organization for Nuclear Research 1211 Gen\`eve 23, Switzerland, CERN}
\author{A. M. Ricci}
\affiliation{INFN Sezione di Pisa, Pisa, Italy}
\author{G. Riccobene}
\affiliation{INFN LNS, Catania, Italy}
\author{E. Richards}
\affiliation{University of Pittsburgh, Pittsburgh, PA 15260, USA}
\author{M. Rosenberg}
\affiliation{Tufts University, Medford, MA 02155, USA}
\author{M. Rossella}
\affiliation{INFN Sezione di Pavia and University, Pavia, Italy}
\author{ N. Rowe}
\affiliation{University of Chicago, Chicago, IL 60637, USA}
\author{P. Roy}
\affiliation{Virginia Tech, Blacksburg, VA 24060, USA}
\author{C. Rubbia}
\affiliation{INFN GSSI, L'Aquila, Italy}
\author{M. Saad}
\affiliation{University of Pittsburgh, Pittsburgh, PA 15260, USA}
\author{I. Safa}
\affiliation{Columbia University, New York, NY 10027, USA}
\author{S. Saha}
\affiliation{University of Pittsburgh, Pittsburgh, PA 15260, USA}
\author{P. Sala}
\affiliation{INFN Sezione di Milano, Milano, Italy}
\author{G. Salmoria}
\affiliation{CBPF, Centro Brasileiro de Pesquisas F\'isicas, Rio de Janeiro, RJ 22290-180, Brazil}
\author{S. Samanta}
\affiliation{INFN Sezione di Genova and University, Genova, Italy}
\author{P. Sapienza}
\affiliation{INFN LNS, Catania, Italy}
\author{A. Scaramelli}
\affiliation{INFN Sezione di Pavia and University, Pavia, Italy}
\author{A. Scarpelli}
\affiliation{Brookhaven National Laboratory, Upton, NY 11973, USA}
\author{D. Schmitz}
\affiliation{University of Chicago, Chicago, IL 60637, USA}
\author{A. Schukraft}
\affiliation{Fermi National Accelerator Laboratory, Batavia, IL 60510, USA}
\author{D. Senadheera}
\affiliation{University of Pittsburgh, Pittsburgh, PA 15260, USA}
\author{S-H. Seo}
\affiliation{Fermi National Accelerator Laboratory, Batavia, IL 60510, USA}
\author{F. Sergiampietri}
\altaffiliation[Now at ]{IPSI-INAF Torino, Italy}
\affiliation{CERN, European Organization for Nuclear Research 1211 Gen\`eve 23, Switzerland, CERN}
\author{G. Sirri}
\affiliation{INFN sezione di Bologna University, Bologna, Italy}
\author{J. S. Smedley}
\affiliation{University of Rochester, Rochester, NY 14627, USA}
\author{J. Smith}
\affiliation{Brookhaven National Laboratory, Upton, NY 11973, USA}
\author{L. Stanco}
\affiliation{INFN Sezione di Padova and University, Padova, Italy}
\author{J. Stewart}
\affiliation{Brookhaven National Laboratory, Upton, NY 11973, USA}
\author{H. A. Tanaka}
\affiliation{SLAC National Accelerator Laboratory, Menlo Park, CA 94025, USA}
\author{F. Tapia}
\affiliation{University of Texas at Arlington, Arlington, TX 76019, USA}
\author{M. Tenti}
\affiliation{INFN sezione di Bologna University, Bologna, Italy}
\author{K. Terao}
\affiliation{SLAC National Accelerator Laboratory, Menlo Park, CA 94025, USA}
\author{F. Terranova}
\affiliation{INFN Sezione di Milano Bicocca, Milano, Italy}
\author{V. Togo}
\affiliation{INFN sezione di Bologna University, Bologna, Italy}
\author{D. Torretta}
\affiliation{Fermi National Accelerator Laboratory, Batavia, IL 60510, USA}
\author{M. Torti}
\affiliation{INFN Sezione di Milano Bicocca, Milano, Italy}
\author{F. Tortorici}
\affiliation{INFN Sezione di Catania and University, Catania, Italy}
\author{R. Triozzi}
\affiliation{INFN Sezione di Padova and University, Padova, Italy}
\author{Y.-T. Tsai}
\affiliation{SLAC National Accelerator Laboratory, Menlo Park, CA 94025, USA}
\author{S. Tufanli}
\affiliation{CERN, European Organization for Nuclear Research 1211 Gen\`eve 23, Switzerland, CERN}
\author{T. Usher}
\affiliation{SLAC National Accelerator Laboratory, Menlo Park, CA 94025, USA}
\author{F. Varanini}
\affiliation{INFN Sezione di Padova and University, Padova, Italy}
\author{S. Ventura}
\affiliation{INFN Sezione di Padova and University, Padova, Italy}
\author{M. Vicenzi}
\affiliation{Brookhaven National Laboratory, Upton, NY 11973, USA}
\author{C. Vignoli}
\affiliation{INFN LNGS, Assergi,  Italy}
\author{B. Viren}
\affiliation{Brookhaven National Laboratory, Upton, NY 11973, USA}
\author{F.A. Wieler}
\affiliation{CBPF, Centro Brasileiro de Pesquisas F\'isicas, Rio de Janeiro, RJ 22290-180, Brazil}
\author{Z. Williams}
\affiliation{University of Texas at Arlington, Arlington, TX 76019, USA}
\author{R. J. Wilson}
\affiliation{Colorado State University, Fort Collins, CO 80523, USA}
\author{P. Wilson}
\affiliation{Fermi National Accelerator Laboratory, Batavia, IL 60510, USA}
\author{J. Wolfs}
\affiliation{University of Rochester, Rochester, NY 14627, USA}
\author{T. Wongjirad}
\affiliation{Tufts University, Medford, MA 02155, USA}
\author{A. Wood}
\affiliation{University of Houston, Houston, TX 77204, USA}
\author{E. Worcester}
\affiliation{Brookhaven National Laboratory, Upton, NY 11973, USA}
\author{M. Worcester}
\affiliation{Brookhaven National Laboratory, Upton, NY 11973, USA}
\author{M. Wospakrik}
\affiliation{Fermi National Accelerator Laboratory, Batavia, IL 60510, USA}
\author{S. Yadav}
\affiliation{University of Texas at Arlington, Arlington, TX 76019, USA}
\author{H. Yu}
\affiliation{Brookhaven National Laboratory, Upton, NY 11973, USA}
\author{J. Yu}
\affiliation{University of Texas at Arlington, Arlington, TX 76019, USA}
\author{A. Zani}
\affiliation{INFN Sezione di Milano, Milano, Italy}
\author{J. Zennamo}
\affiliation{Fermi National Accelerator Laboratory, Batavia, IL 60510, USA}
\author{J. Zettlemoyer}
\affiliation{Fermi National Accelerator Laboratory, Batavia, IL 60510, USA}
\author{C. Zhang}
\affiliation{Brookhaven National Laboratory, Upton, NY 11973, USA}
\author{S. Zucchelli}
\affiliation{INFN sezione di Bologna University, Bologna, Italy}

%% file: abstract.tex
We present a search for long-lived particles (LLPs) produced from kaon decay that decay to two muons inside the ICARUS neutrino detector. This channel would be a signal of hidden sector models that can address outstanding issues in particle physics such as the strong CP problem and the microphysical origin of dark matter. The search is performed with data collected in the Neutrinos at the Main Injector (NuMI) beam  at Fermilab corresponding to $2.41\times 10^{20}$ protons-on-target.
No new physics signal is observed, and we set world-leading limits on heavy QCD axions, as well as for the Higgs portal scalar among dedicated searches. Limits are also presented in a model-independent way applicable to any new physics model predicting the process $K\to \pi+S(\to\mu\mu)$, for a long-lived particle S. This result is the first search for new physics performed with the ICARUS detector at Fermilab. It paves the way for the future program of long-lived particle searches at ICARUS.

%% file: Introduction.tex
Several beyond standard model (BSM) physics models predict processes by which a kaon ($K^{\pm}$ or $K^0_L$) decays to a long-lived particle (LLP), which in turn decays to a $\mu^+ \mu^-$ pair. Two such processes are the Higgs portal scalar (HPS)~\cite{HPSSBN}, a dark sector model for dark matter~\cite{HPS-DM}, and a heavy QCD axion, or axion-like particle (ALP)~\cite{uALP, KALP}, an $\mathcal{O}(\mathrm{GeV})$ resolution to the strong-CP problem~\cite{CP-Problem}. Under these scenarios kaons created in the target of the Neutrinos at the Main Injector (NuMI) beam at Fermilab would produce (psuedo-)scalars that could then propagate into the ICARUS detector and decay into di-muon pairs. ICARUS sits \SI{800}{m} downstream of the NuMI target at a 5.75$^\circ$ far-off-axis position.

The ICARUS liquid argon time projection chamber (LArTPC) neutrino detector has been operational at Fermilab since 2022~\cite{ICARUSOG, ICARUSOverhaul, ICARUSElectronics, ICARUSInaguralPaper}, taking data as part of the Short-Baseline Neutrino (SBN) Program~\cite{SBNProposal, SBNProgram}. 
ICARUS consists of two cryostat modules both containing two TPCs with a common cathode. Each constituent TPC identifies neutrino interactions through the detection of ionization charge deposited in tracks and electromagnetic showers by charged particles produced in neutrino-argon interactions. The ionization charge is used to reconstruct charged particle trajectories with good calorimetry and precise ($\sim$\si{mm}) spatial resolution.

Calorimetric, topological, and kinematic features distinguish di-muon decays in ICARUS from neutrino backgrounds. The primary residual neutrino background, from muon neutrino charged current coherent pion production ($\uCCCoh$)~\cite{ReinSehgal, BergerSehgal}, is tuned from external data and further fit to a sideband in the analysis. The analysis searches for di-muon decays in a fiducial volume where both muons are contained in the detector. An excess above the expected neutrino background in a narrow region of invariant mass consistent with a resonance peak would be a signature of new physics. The results are interpreted through the Higgs portal scalar and heavy axion benchmark models. A model independent interpretation  applicable to any model predicting the process $K \to \pi + S (\to \mu\mu)$ is also included.

%% file: Models.tex
\section{LLP Models}
\textit{Higgs Portal Scalar: } \input{HPS}

\textit{Heavy QCD Axion: }\input{ALP}

%% file: HPS.tex
The Higgs portal scalar extends the standard model (SM) with a real, neutral scalar $S$ with mass $m_S$ that mixes with the SM Higgs boson with a mixing angle $\theta_S$. Such a scalar is a candidate portal between the standard model and a dark matter particle that would enable dark matter to be thermally produced in the early universe~\cite{HPS-DM}, while preventing the overabundance that would result if the dark matter particle were light (its mass less than a few GeV) and coupled to the electroweak bosons~\cite{LDM-Limit}. In the mass region of interest ($2m_\mu < m_S < m_K - m_\pi$), the scalar is produced in decays of charged ($K^\pm$) and neutral ($K^0_L$) kaons and predominantly decays to pairs of muons and pions~\cite{HPSSBN}.

%% file: ALP.tex
The axion is a solution to the strong-CP problem, the experimental observation that the CP-violating QCD coupling is very small ($ \lesssim 10^{-10}$) despite no symmetry in the standard model requiring it to be at that scale~\cite{CP-Problem}. The traditional QCD axion model implies the existence of a light axion particle with a very large decay constant~\cite{PQ1, PQ2}, and suffers from the quality problem~\cite{QP1, QP2}. Extended axion models, such as those incorporating a larger gauge group~\cite{Dimopoulos:1979pp,Rubakov:1997vp,Gherghetta:2016fhp,Dimopoulos:2016lvn,Agrawal:2017ksf,Agrawal:2017evu,Gaillard:2018xgk,Lillard:2018fdt,Csaki:2019vte,Gherghetta:2020keg,Gherghetta:2020ofz,Valenti:2022tsc}, or a mirror symmetry~\cite{Berezhiani:2000gh,Hook:2014cda,Hook:2019qoh,Fukuda:2015ana}, avoid the quality problem by engineering a heavier axion particle with a smaller decay constant. 

Such an axion-like particle (ALP) can be introduced through a low-energy effective Lagrangian independent of the UV physics
\begin{equation*}
    \mathcal{L} \supset \frac{c_3 \alpha_s a}{8\pi f_a}G\Tilde{G} + \frac{c_2\alpha_2 a}{8\pi f_a} W\Tilde{W} +  \frac{c_1\alpha_1 a}{8\pi f_a} B\Tilde{B} +  \frac{c_\mu\partial_\nu a}{2 f_a}\Bar{\mu}\gamma^\nu\gamma_5\mu\,,
\end{equation*}
where $f_a$ is the axion decay constant and $\alpha_s, \alpha_2, \alpha_1$ are given by the Standard Model gauge couplings ($\alpha_i = g_i^2/4\pi$) to the gluons ($G$), $SU(2)$ gauge field ($W$), and $U(1)$ gauge field ($B$), respectively~\cite{uALP, KALP}. The axion couples to each gauge field through the coupling $c_i$, as well as to the muon by a coupling $c_\mu$. The muon coupling can be included at tree level~\cite{uALP}, or induced indirectly from the renormalization group flow from the axion decay constant scale ($f_a$) to the mass scale ($m_a$)~\cite{KALP, Chala:2020wvs, Bauer:2020jbp}. For this result, we consider the popular ``co-dominance'' model where there is no tree-level muon coupling and the gauge boson couplings are equal: $c_1 = c_2 = c_3 = 1$. 

In this case, axions are produced in charged kaon decay~\cite{AxionStrongProduction, AxionStrongProductionDetailed, AxionStrongProduction2ndOrder}. Production in neutral kaon decay is suppressed: for $K^0_L$ by CP-violation, and for $K^0_S$ by the short lifetime of the particle. The axions decay predominantly to $\gamma\gamma$ and $\mu\mu$ final states. The $ee$ decay is suppressed by the small mass of the electron, while (semi-)hadronic axion decays are not significant for $m_a \lesssim$~\SI{0.4}{GeV}~\cite{Aloni:2018vki}, in the mass region of interest. The di-gamma decay depends on the effective axion-photon coupling ($c_\gamma$), for which we apply a computation based on vector meson dominance~\cite{KALP, Aloni:2018vki}.

%% file: SignalBox.tex
A Monte Carlo simulation developed for ICARUS includes models of the NuMI neutrino and cosmic backgrounds, as well as the scalar signal. The cosmic ray flux is simulated with the CORSIKA generator~\cite{Corsika}. The NuMI flux is simulated with g4numi~\cite{g4numi}, a GEANT4~\cite{GEANT4} based hadron production and participle propagation simulation of NuMI proton beam on the full NuMI beamline geometry. Hadron production data from NA49~\cite{NA49A, NA49B} are used to correct hadron production cross sections~\cite{NuMIFlux}. Neutrino interactions are simulated by the GENIE framework (v3.0.6  Ar23.20i)~\cite{GenieUserManual, GENIE}. The dominant background component in the signal region consists of muon neutrino charged current coherent pion production. We have tuned the GENIE implementation of the Berger-Sehgal $\uCCCoh$ model~\cite{BergerSehgal} to a measurement of the process performed by MINERvA~\cite{MINERvA} (see appendix). Energy depositions from particles traversing the detector are simulated using GEANT4~\cite{GEANT4}. 
The response of the ICARUS TPC is simulated by the Wire-Cell framework~\cite{WireCellMC}, with the ionization signal shapes tuned to ICARUS data~\cite{ICARUSSignalNoise}. The HPS~\cite{HPSSBN} and ALP~\cite{KALP} signal models are simulated based on the kaon flux extracted from the NuMI $\nu$ flux simulation. The scalar flux peaks at $E_S\sim\,$\SI{0.5}{GeV} with a broad tail out to a few GeV.

Ionization depositions in the detector recorded by the TPC are reconstructed into “events”. The reconstruction is performed by the Pandora framework~\cite{Pandora, DUNEPandora}, which groups ionization hits into events consisting of track-like and shower-like objects and an interaction vertex. This analysis identifies candidate di-muon decays as Pandora events with two tracks, where the track start point is inside the fiducial volume and both tracks stop inside a containment volume inside the detector. 
The fiducial volume is defined with an inset of \SI{10}{cm} from the active volume in the vertical and drift ($\hat{x}$ and $\hat{y}$) directions, \SI{15}{cm} at the front in the beamline ($\hat{z}$) direction, and \SI{1}{m} at the back. 
The containment volume is the same except it is at an inset of \SI{15}{cm} at the back of $\hat{z}$. Problematic regions of the detector are also removed from both volumes.


Candidate track pairs are subjected to topological and calorimetric cuts which require consistency with muon-like energy deposition patterns. First, the muon momentum reconstructed by the stochastic multiple-Coulomb-scattering (MCS) along the track~\cite{MicroBooNEMCS} is compared to the muon momentum reconstructed by the track range~\cite{PDG}. The track range identifies the momentum with a resolution of $\sim 3$\%, whereas MCS obtains a resolution of 8-20\% depending on the track length.
A cut is made such that the momentum determined by MCS is not greater than 50\% above the range-based momentum. This cut rejects proton tracks, which are typically straighter than muons and are reconstructed with a much larger MCS momentum than their track range would indicate.

Second, calibrated ionization energy depositions reconstructed along the last \SI{25}{\centi\meter} of the track are compared to the muon and proton expectation to build $\mu$-like and $p$-like particle identification (PID) scores~\cite{ICARUSEnergyScale, ArgoNeuTPID}. Muon energy loss in this track range forms a Bragg peak, which distinguishes stopping muons from protons and interacting pions. The PID cuts accept 76\% of scalar-induced muons while rejecting 55\% of neutrino-induced pions and protons. 

A signature of di-muon decays is that, unlike neutrinos, there is no nuclear target, and thus no source for hadronic (highly-ionizing) energy depositions below the tracking threshold or additional charge deposits around the vertex.
To leverage this, cuts are applied on large charge depositions close to the candidate decay vertex. 
These cuts reject protons in an energy range $15 \lesssim \text{proton K.E.} \lesssim$~\SI{50}{MeV}, below the current Pandora tracking threshold in ICARUS.

Kinematic cuts further distinguish di-muon decay events. The momenta of the predicted scalars is such that the di-muon products have a small opening angle. A cut is made at $70^\circ$ on the reconstructed di-muon opening angle. Kaons generally decay near the NuMI target, thus the momentum vector for most scalars points along the direction from the NuMI target to the ICARUS detector. The angle between these two vectors ($\theta_\text{NuMI}$) is reconstructed as the angle between the summed di-muon momentum vector (obtained from the track direction and range) and the known direction from the NuMI target to the ICARUS detector. The pointing resolution on this angle for scalar decays is $\sim 2^\circ$, while the distribution for neutrino interactions is much more broad and peaks at about 20$^\circ$. The full $\theta_\text{NuMI}$ range is used in the analysis. Events where $\theta_\text{NuMI}>5^\circ$ are used to characterize the background, while the $\theta_\text{NuMI}<5^\circ$ region defines the final signal selection.

The signal selection identifies signal scalar events with an efficiency of 9-18\% (depending on the scalar mass). All cosmic activity and 99.96\% of beam induced neutrino interactions within the fiducial volume are rejected in a simulated sample equivalent to roughly 100$\times$ data exposure. The di-muon invariant mass is reconstructed with a resolution of 2-5\%, depending on the scalar mass. The binning of the di-muon mass spectrum is optimized to match this resolution.

%% file: Systematics.tex
Systematic uncertainties from the models used in the Monte Carlo simulation (flux, interaction cross sections, particle propagation, and detector response) influence the predicted event rates for signal and background channels. The size of their impact is summarized in Table \ref{tab:syst}. These uncertainties, in combination with the simulation, have been validated by comparisons of data and simulation in sideband regions in the analysis. 

The neutrino flux uncertainty is estimated from hadron production data~\cite{NA49A, NA49B}, assumed to be 40\% outside of data coverage. Focusing and operational uncertainties associated with the NuMI beamline are estimated by simulating fluxes with alternate geometry configurations~\cite{NuMIFlux}.
An additional flat normalization uncertainty of 8.6\% added in quadrature to account for uncertainties in the extended NuMI target hall geometry that are particularly relevant for the ICARUS far-off-axis location. The flux uncertainty for kaon-induced scalars is computed from the components of the neutrino flux uncertainty relevant for kaon production. The flux uncertainty is treated as correlated between the neutrino background and scalar signal. 

The uncertainty on the $\uCCCoh$ cross section, which is the dominant background uncertainty, is derived from a tuning procedure detailed in the appendix. The cross section uncertainty from all other processes is obtained from GENIE~\cite{GenieUserManual}. Uncertainties on particle propagation account for uncertainties in $\pi$-Ar and $p$-Ar interaction cross sections and are computed by the GEANT4-Reweight package~\cite{Geant4Reweight}. They are relevant only for the neutrino background.

\begin{table}
    \centering
    \begin{tabular}{l | p{1.5cm} | p{1.5cm}}
\rowcolor{gray!10}
        Systematic & Scalar Sig. [\%] & Neutrino Bkg. [\%] \\
    \hline
    Total Detector Uncertainty & 11.0 & 20.2\\
    \quad Detector Model Variations & 9.9 & 17.6\\
    \quad Cathode Aplanarity & 5.5 & 5.6 \\
    \quad Energy Scale & 1.8 & 8.2\\
    Flux & 12.3 & 12.0 \\
    $\uCCCoh$ x-sec & -- & 62.9\\
    Other $\nu$ Interactions x-sec & -- & 4.1\\
    Particle Propagation & -- & 5.6\\
    MC. Stat & -- & 4.6\\
    \hline
    \rowcolor{gray!10}
    Total & 16.5 & 67.7 \\
    \end{tabular}
    \caption{Uncertainty on the total event rate in the signal region for scalar signals and neutrino backgrounds. The scalar signal uncertainty is taken as the mean across the sampled mass points. It is nearly independent of the mass.}
    \label{tab:syst}
\end{table}

Uncertainties on the detector response arise from three sources: detector model variations, reconstructed energy scales, and cathode aplanarity. Aspects of the detector model such as the effective channel gain, noise level, and ionization signal shape are varied by the size of the variation of each quantity across the runtime of the ICARUS experiment. 
Separate simulations with each of the detector model variations are performed, and the change in the number of events entering the signal region is propagated as a normalization uncertainty on the rate. 
The energy scale uncertainties are relevant for the two reconstructed energy scales used in the event selection: the MCS momentum and the calorimetric ionization energy. The uncertainty is implemented by varying the reconstructed energies in Monte Carlo simulation by an amount prescribed by the calibration. Finally, there is an uncertainty to address the aplanarity of the cathode plane in both ICARUS cryostats. 

%% file: SignalBoxStat.tex
The analysis of di-muon-like events is performed as a search for an excess above background or “bump hunt” within the reconstructed di-muon invariant mass spectrum. 
It makes use of the BumpHunter test statistic \cite{Choudalakis:2011qn, pyBumpHunter}, which is defined as $-\log p_\text{min}$, where $p_\text{min}$ is the minimum local p-value among a set of windows over the binned di-muon invariant mass spectrum.

The signal region search is performed in three steps. First, the di-muon mass window with the greatest excess above the nominal background in the signal region ($\theta_\text{NuMI} < 5^\circ$) is identified by the BumpHunter algorithm. Second, a scale factor on the $\uCCCoh$ background template is fit to data in a larger $\theta_\text{NuMI}$ region ($\theta_\text{NuMI} < 10^\circ$), excluding the invariant mass range identified in the first step. Third, the BumpHunter algorithm is rerun to identify the largest excess in the signal region above the scaled background. The global significance of the BumpHunter test statistic output from the final step is obtained using the null hypothesis BumpHunter test statistic distribution as computed with 10,000 toy experiments. In each toy experiment, systematic and statistical uncertainties on the neutrino rate are thrown and the full three-step procedure is run to obtain the sampled  test statistic value. 

%% file: Results.tex
\begin{figure}[t]
    \centering
    \includegraphics[width=\linewidth]{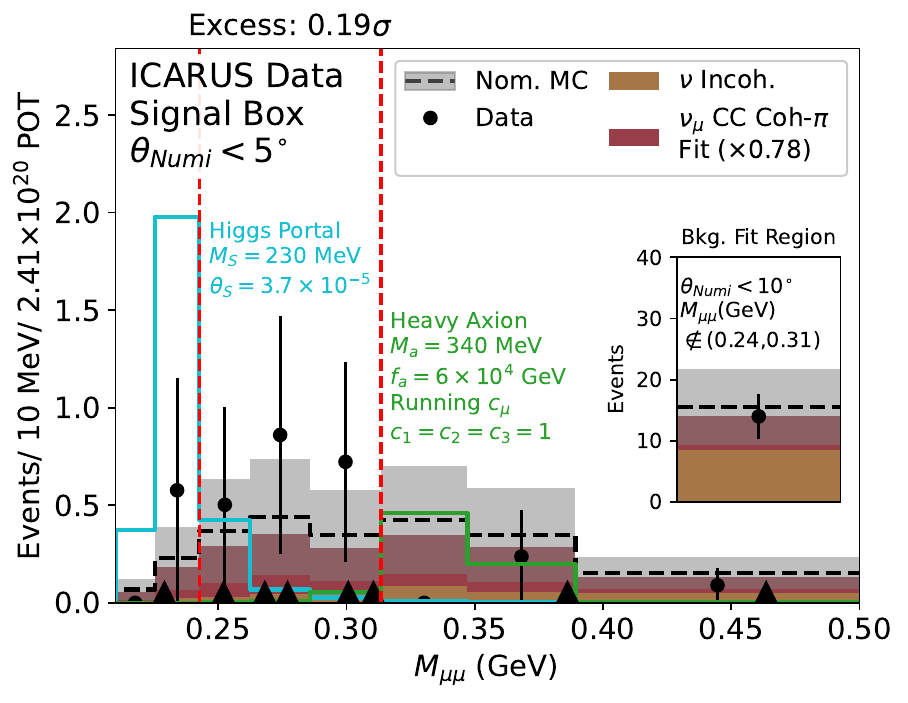}
    \caption{Signal box result of the search. A scale factor on the $\uCCCoh$ background is obtained from a fit region shown in the inset. A bump hunt search in the data against the scaled background prediction obtains an insignificant excess with a global significance of $0.19\sigma$, in the range $0.24 < M_{\mu\mu} < $~\SI{0.31}{GeV}. The reconstructed mass of the eight data events are shown with a caret on the bottom of the figure. Spectra of two example model points excluded at the 90\% CL (see Fig.~\ref{fig:ModelLim}) are shown.}
    \label{fig:Result}
\end{figure}

The search is performed with data corresponding to $2.41 \times 10^{20}$ protons-on-target (POT) with the NuMI beam in the Forward Horn Current (FHC) configuration, taken between June 2022 and July 2023. Beam quality cuts, based on those developed by NO$\nu$A~\cite{NOvABeam}, are applied to the data.
The result of the search is shown in Fig.~\ref{fig:Result}. A scale factor of 0.78 on the $\uCCCoh$ component of the background is obtained from the template fit. The initial background expectation in the signal box region is $7.8 \pm 5.0~(\text{syst.}) \pm 2.8~(\text{stat.})$, fit to $6.4$ events by the scale factor. Eight data events are observed. The largest excess is found in the mass window $0.24 < M_{\mu\mu} < $~\SI{0.31}{GeV}, with a global significance of $0.19\sigma$ when compared to the null hypothesis. The excess is not statistically significant, and limits are set with the CL$_\text{s}$ method at the 90\% confidence level (CL)~\cite{CLs}. The obtained scale factor is consistent within the prior uncertainty on the $\uCCCoh$ rate (62.9\%), especially noting that it is subject to a significant statistical uncertainty, and also biased downward by the bump estimate procedure.


\begin{figure}[h]
    \centering
    \includegraphics[width=0.9\linewidth]{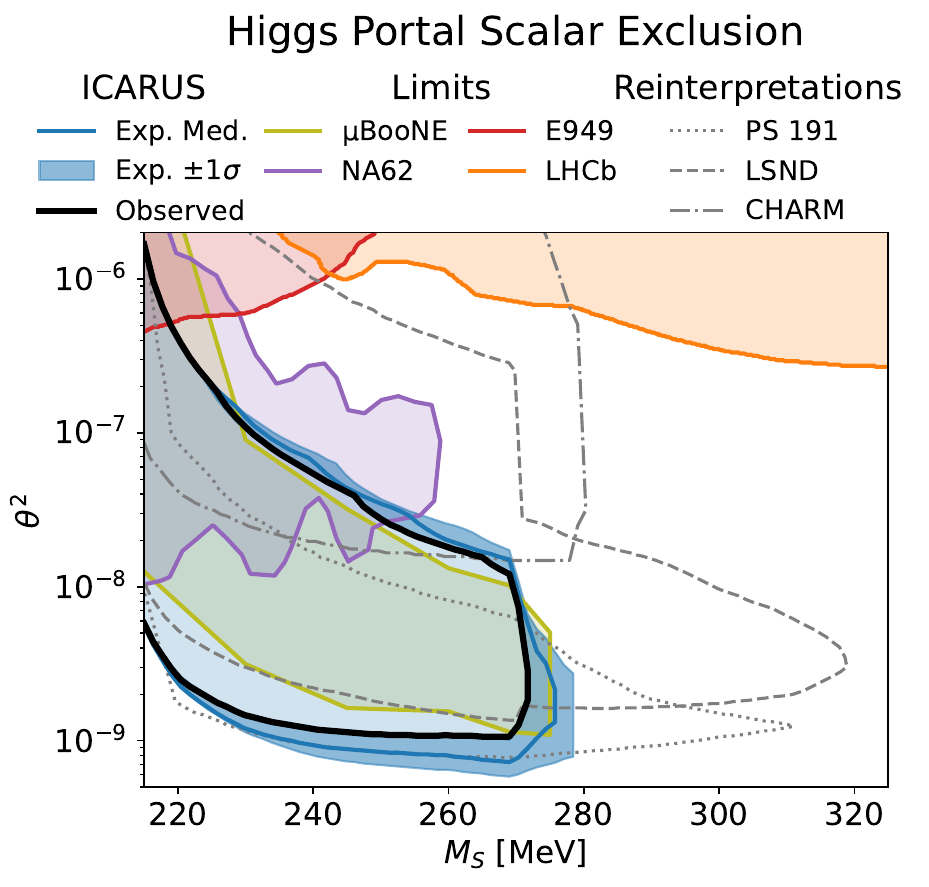}
    \includegraphics[width=0.9\linewidth]{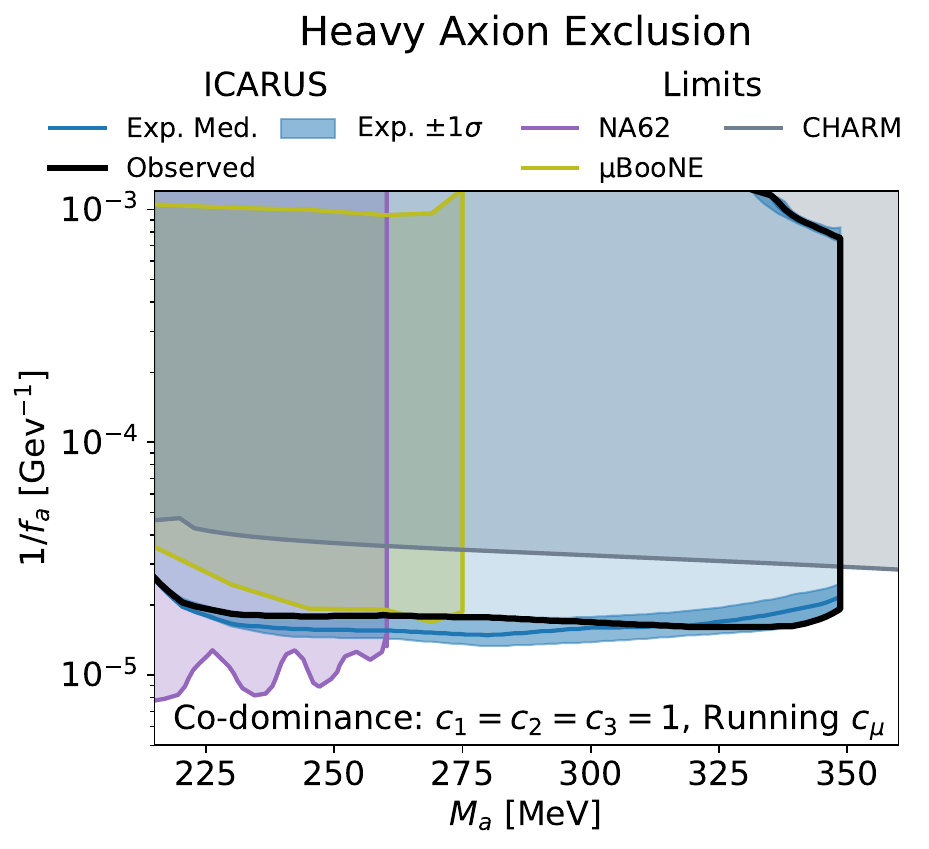}
    \caption{Limits on the Higgs Portal scalar (top) and heavy axion (bottom) models. Exclusions are computed with the CL$_\text{s}$ method at the 90\% CL. The HPS limit is compared to direct searches~\cite{NA62HPS, MicroBooNEHPS, LHCbHPS1, LHCbHPS2, E949} and reinterpreted limits on the model~\cite{CHARMInterp, CHARMResult, PS191Interp, PS191Result, LSNDInterp, LSNDResult}. The ALP limit is compared to other limits on the particle~\cite{KALP, NA62HPS, MicroBooNEHPS, ALPINIST, CHARMResult}.}
    \label{fig:ModelLim}
\end{figure}

Limits on the Higgs portal scalar and heavy QCD axion models are shown in Fig.~\ref{fig:ModelLim}. This work obtains world leading limits on the heavy axion decay constant, covering lower values of $1/f_a$ in the axion mass range \SI{260}{MeV}~$\lesssim M_a < m_K - m_\pi$.
The limits from this search also extend to larger (tree-level) axion-muon couplings up to $c_\mu \approx 10^{-4}$--$10^{-1}$ and are complementary to prior searches for muon-coupled axions~\cite{uALP,ArgoNeuTAxion, CHARMResult, NA48, LHCbHPS1}. This measurement also extends the limits on the Higgs portal scalar mixing angle amongst dedicated searches in the mass range $2m_\mu < m_S \lesssim$~\SI{270}{MeV}. Reinterpretations of measurements from LSND~\cite{LSNDInterp, LSNDResult} and PS191~\cite{PS191Interp, PS191Result} extend further, but are based on re-analyses of other measurements performed outside of both experiments. The observed exclusions for both models are consistent with the range of expected sensitivity.

\begin{figure}[t]
    \centering
    \includegraphics[width=\linewidth]{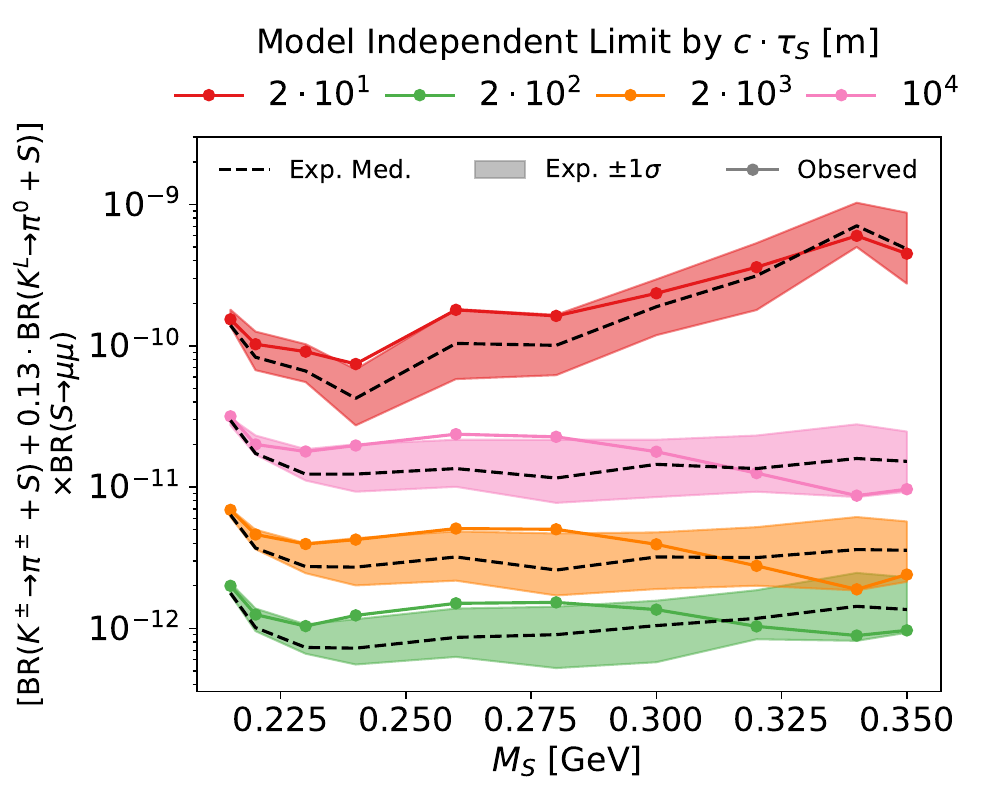}
    \caption{Model independent limits (at the 90\% CL) on the process $K\to \pi + S(\to\mu\mu)$, for a long lived particle $S$. Contours are shown as a function of the branching ratio of the process and the particle mass, for a few example values of the particle lifetime. Limits in the full three-dimensional space are available in the supplementary material.}
    \label{fig:ModelIndepLim}
\end{figure}

Model independent limits are shown in Fig.~\ref{fig:ModelIndepLim}. The model independent scenario interprets the reach of the search for a general branching ratio of the process, scalar lifetime $\tau_S$, and scalar mass $M_S$. 
Both charged $K^{\pm}$ and $K^0_L$ contribute to scalar production.
They produce scalars with essentially the same energy spectrum, and do so at a relative rate, $1K^\pm:0.13K^0_L$, independent over the sampled scalar mass points.
Thus, the combined branching ratio 
$(\text{BR}(K^\pm\to \pi^\pm S) + 0.13\cdot\text{BR}(K^0_L\to \pi^0 S))\times \text{BR}(S\to\mu\mu)$
is a model independent parameter that defines the reach of the search. In Fig.~\ref{fig:ModelIndepLim}, the combined branching ratio is plotted for a few choices of the scalar lifetime. The limits in the full three dimensional model independent space are available as supplementary data.

%% file: Conclusion.tex
We have presented a search for a di-muon resonance produced in kaon decay performed with the ICARUS neutrino detector with $2.41\times 10^{20}$ POT in the NuMI beam in its FHC configuration. The search finds no significant excess above the expected neutrino background. World leading limits are set on heavy QCD axions for various choices of the muon and gauge boson couplings including the co-dominance scenario, and leading limits among dedicated searches are set for the Higgs Portal scalar in the mass range $2m_\mu < m_S \lesssim$~\SI{270}{MeV}. The search is also interpreted in a model independent framework applicable to any new physics model predicting the process $K \to \pi + S (\to \mu\mu)$, for a long-lived scalar $S$.

This search is the first result with the ICARUS neutrino detector at Fermilab. It paves the way to a future program of hidden sector searches at ICARUS, as well as the broader set of measurements planned as part of the Short-Baseline Neutrino program.

%% file: acknowledgements.tex
This document was prepared by the ICARUS Collaboration using the resources of the Fermi National Accelerator Laboratory (Fermilab), a U.S. Department of Energy, Office of Science, HEP User Facility. Fermilab is managed by Fermi Research Alliance, LLC (FRA), acting under Contract No. DE-AC02-07CH11359. This work was supported by the US Department of Energy, INFN, EU Horizon 2020 Research and Innovation Program under the Marie Sklodowska-Curie Grant Agreement No. 734303, 822185, 858199 and 101003460, and the Horizon Europe Research and Innovation Program under the Marie Sklodowska-Curie Grant Agreement No. 101081478 Part of the work resulted from the implementation of the research Project No. 2019/33/N/ST2/02874 funded by the National Science Centre, Poland. We also acknowledge the contribution of many SBND colleagues, in particular for the development of a number of simulation, reconstruction and analysis tools which are shared within the SBN program.

%% file: cohpimodel.tex
The Rein-Sehgal model \cite{ReinSehgal} and the Berger-Sehgal model \cite{BergerSehgal} are based on Adler's PCAC relation, which predicts the $\uCCCoh$ cross section to be proportional to the pion-nucleus ($\pi-N$) elastic scattering cross section as
\begin{equation}\label{eq:xsec}
\begin{split}
    \frac{d\sigma^{CC}}{dQ^2dydt} = {} & \frac{G_F^2 cos^2\theta_C f^2_{\pi}}{2\pi^2} \frac{E}{|\boldsymbol{q}|^2}uv \\
    & \hspace{-45pt} \times \left[ \left(G_A- \frac{1}{2} \frac{Q_{min}^2}{Q^2+m_{\pi}^2}\right)^2 \right. \left. +\frac{y}{4} (Q^2-Q^2_{min}) \frac{Q^2_{min}}{(Q^2+m^2_{\pi})^2} \right]\\
    & \hspace{-45pt} \times \frac{d\sigma(\pi^+ N \rightarrow \pi^+ N)}{dt},
\end{split}
\end{equation}
where $t$ is the momentum transfer to the nucleus, $Q^2$ is the 4-momentum transfer squared, $Q_{min}^2 \, = \, m_l^2 y / (1-y)$ is the high energy approximation to the true minimal $Q^2$, $y$ is the Bjorken inelasticity, and $G_A$ is the axial vector form factor. The kinematic factors $u$ and $v$ are $u,v \, = \, (E + E' \pm |\bold{q}|)/2E$. Constants $G_F$, $\theta_C$ and $f_{\pi}$ are the Fermi coupling constant, Cabibbo angle, and the pion decay constant, respectively. The two models differ in the way they calculate the $\pi-N$ differential cross section. The Rein-Sehgal model derives it from the pion-nucleon ($\pi-n$) cross section using the ansatz
\begin{equation}\label{eq:RSxesc}
    \frac{d\sigma(\pi N \rightarrow \pi N)}{dt} = A^2 \frac{d \sigma_{el}}{dt}\Bigr|_{t=0} e^{-bt} F_{abs},
\end{equation}
where $b = \frac{1}{3}R_0^2 A^{2/3}$, $F_{abs} = \mathrm{exp} \left( -\frac{9 A^{1/3}}{16 \pi R_0^2} \sigma_{inel} \right)$, and the forward elastic $\pi-n$ differential cross section is given by the optical theorem as 
\begin{equation}
    \frac{d \sigma_{el}}{dt}\Bigr|_{t=0} = \frac{1}{16\pi}\left( \frac{\sigma_{tot}^{\pi^+ p} + \sigma_{tot}^{\pi^- p}}{2} \right)^2,
\end{equation}
with $\sigma_{tot}^{\pi^{\pm p}}$ calculated from pion-deuterium scattering data.
The Berger-Sehgal model uses a different ansatz for $T_\pi <$~\SI{1}{GeV}
\begin{equation}\label{eq:BSxec}
    \frac{d\sigma(\pi N \rightarrow \pi N)}{dt} = A_1 \left(\frac{A}{12}\right)^{4/3} e^{-b_1 \left(\frac{A}{12}\right)^{2/3} t }, 
\end{equation}
where $A_1$ and $b_1$ are pion energy dependent constants obtained by fitting to pion-carbon scattering data~\cite{PiArData}. Terms containing the nucleus mass number $A$ models the $A$-dependence behavior.

\begin{figure}[t]
    \centering
    \includegraphics[width=\linewidth]{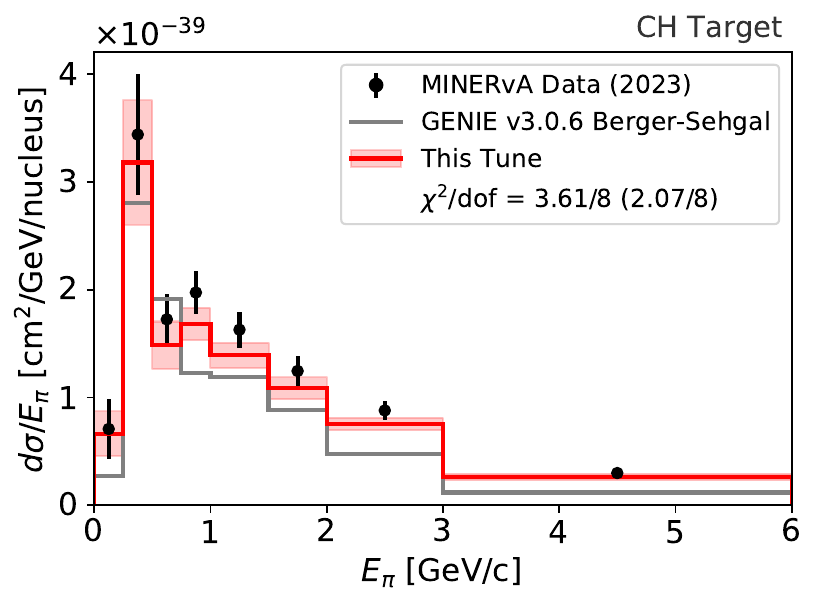}
    \includegraphics[width=\linewidth]{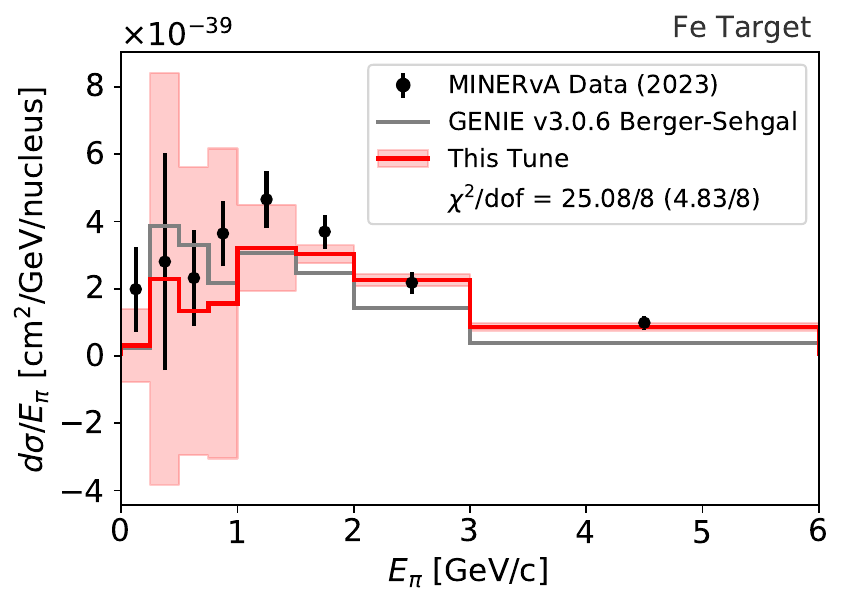}
    \caption{$\uCCCoh$ differential cross section as a function of pion energy predicted by GENIE v3.00.06 and this tune, compared to the MINERvA measurement.}
    \label{fig:CohpiFitEpi}
\end{figure}

We tune the GENIE v3.0.6 $\uCCCoh$ model event generator to MINERvA cross section measurements on the hydrocarbon (CH) and iron ($^{56}$Fe) target. This fit uses the Berger-Seghal framework for the $\uCCCoh$ process, but infers the $\pi$-Ar cross section from the MINERvA measurement.
The tune was performed using the NUISANCE package \cite{NUISANCE} with the GENIE Reweight package, which together provides the infrastructure to fit MC simulation to data by varying model parameters in GENIE and extract fit uncertainties from systematic throws. The fit was performed in two steps: the A-dependence fit and the cross section scaling fit. First, the exponent of the first A-dependence term from equation \ref{eq:BSxec} was varied to fit the measurement of the CH/Fe total cross section ratio as a function of neutrino energy. This A-dependence tune enables realistic extension of the model to the argon nucleus to within the fit uncertainty ($\sim$60\%). Second, eight scaling factors that scale the $\pi-N$ cross section for the eight pion energy bins in Fig.~\ref{fig:CohpiFitEpi} were varied for a joint fit to the differential cross section with respect to pion energy for the CH and $^{56}$Fe targets. 
The fit tunes the Berger-Sehgal model to be based on the MINERvA data in place of the original pion-carbon scattering data, for which the information on measurement uncertainty or the covariance matrix is unavailable.

\begin{table}[t]
    \centering
    \begin{tabular}{l | p{1.2cm} | l}
    \rowcolor{gray!10}
    Model Parameter & Nominal Value & Fit Value \\
    \hline
    A-dependence factor & 4/3 & 0.91$\, \pm \,$0.46 \\
    \hline
    $\sigma(\pi N)$ scales  &  & \\
    \quad $0 \,\,\,\,\,\,\,\, < E_{\pi}\,[\mathrm{GeV}] < 0.25$ & 1 & 2.9 $\, \pm \,$1.1 \\
    \quad $0.25 < E_{\pi}\,[\mathrm{GeV}] < 0.5$        & 1 & 1.1$\, \pm \,$0.2 \\
    \quad $0.5 \,\,\, < E_{\pi}\,[\mathrm{GeV}] < 0.75$ & 1 & 0.7$\, \pm \,$0.1 \\
    \quad $0.75 < E_{\pi}\,[\mathrm{GeV}] < 1.0$        & 1 & 1.3$\, \pm \,$0.1 \\
    \quad $1.0 \,\,\, < E_{\pi}\,[\mathrm{GeV}] < 1.5$  & 1 & 1.1$\, \pm \,$0.1 \\
    \quad $1.5 \,\,\, < E_{\pi}\,[\mathrm{GeV}] < 2.0$  & 1 & 1.2$\, \pm \,$0.1 \\
    \quad $2.0 \,\,\, < E_{\pi}\,[\mathrm{GeV}] < 3.0$  & 1 & 1.5$\, \pm \,$0.1 \\
    \quad $3.0 \,\,\, < E_{\pi}\,[\mathrm{GeV}] < 6.0$  & 1 & 2.2$\, \pm \,$0.3 \\
    \hline
    \end{tabular}
    \caption{Model parameters values and uncertainties from GENIE v3.00.06 and this tune.}
    \label{tab:tunevals}
\end{table}

Fitted parameter values are summarized in Table \ref{tab:tunevals}. A comparison of the tune result with the MINERvA data and GENIE v3.0.6 for the differential cross section with respect to the pion energy is shown in Fig.~\ref{fig:CohpiFitEpi}. $\chi^2/$ndof for $d\sigma/dE_{\pi}$ is 3.61/8 for the CH target and 25.08/8 for the $^{56}$Fe target. With the fit uncertainty taken into account, the $\chi^2/$ndof becomes 2.07/8 for the CH target and 4.83/8 for the $^{56}$Fe target, which shows that the tuned result agrees well with the MINERvA data within the fit uncertainty. 

The result of the fit and its uncertainty is taken as the central value and uncertainty of the $\uCCCoh$ cross section for the search. An additional uncertainty on the axial mass of the form factor of $\pm$\SI{0.3}{GeV} is also added in quadrature. This uncertainty is added to cover the range of predictions for the phenomenological form factor $G_A$ (see, e.g., Ref.~\cite{BergerSehgal, Belkov:1986hn, Paschos1}). Its impact is sub-leading to the uncertainties on the pion-argon cross section.
The large uncertainty on the A-dependence parameter results in large uncertainties for targets other than carbon. In addition, large measurement uncertainties on low pion energy bins results in large fit uncertainties on corresponding cross section scaling factors.  Nonetheless, this tune provides a more reliable uncertainty than the flat 100\% assigned by GENIE, and also includes realistic shape uncertainties on the kinematic variables relevant for $\uCCCoh$ cross section modeling.